\documentstyle[prd,aps,preprint,psfig]{revtex}

 \newcommand{\CL}{{\cal L}}

 \newcommand{\bea}{\begin{eqnarray}}  \newcommand{\eea}{\end{eqnarray}}
 \newcommand{\beq}{\begin{equation}}  \newcommand{\eeq}{\end{equation}}
   
 \newcommand{\lmk}{\left(}  \newcommand{\rmk}{\right)}
 \newcommand{\lkk}{\left[}  \newcommand{\rkk}{\right]}
   
 \newcommand{\del}{\partial}  
 
 \newcommand{\bib}{\bibitem} 
 \newcommand{\la}{\left\langle} \newcommand{\ra}{\right\rangle}
  
 \newcommand{\gtilde} {~ \raisebox{-1ex}{$\stackrel{\textstyle >}{\sim}$} ~}

\def\IBD#1#2#3{{\it ibid}. D{\bf #1}, #2 (19#3)}
\def\NPB#1#2#3{Nucl. Phys. {\bf B#1}, #2 (19#3)}

\def\PLB#1#2#3{Phys. Lett. B{\bf #1}, #2 (19#3)}
\def\PLBold#1#2#3{Phys. Lett. {\bf#1B}, #2 (19#3)}

\def\PRD#1#2#3{Phys. Rev. D{\bf #1}, #2 (19#3)}

\def\PRL#1#2#3{Phys. Rev. Lett. {\bf#1}, #2 (19#3)}
\def\PRT#1#2#3{Phys. Rep. {\bf#1}, #2 (19#3)}

\def\PTP#1#2#3{Prog. Theor. Phys. {\bf #1}, #2 (19#3)}

\def\MNRAS#1#2#3{Mon. Not. R. Astron. Soc. {\bf #1}, #2 (19#3)}

\begin{document}
\draft
\preprint{\begin{tabular}{l}
\hbox to\hsize{\hfill UT-854}\\
\hbox to\hsize{\hfill RESCEU-20/99}\\
\hbox to\hsize{\hfill UTAP-332}\\
\end{tabular}}
\title{Initial condition for new inflation in supergravity}
\author{T. Asaka}
\address{\it Department of Physics, School of Science, University of Tokyo, Tokyo~113-0033, Japan}
\author{M. Kawasaki}
\address{\it Research Center for the Early Universe, School of
  Science, University of Tokyo, Tokyo~113-0033, Japan}
\author{Masahide Yamaguchi}
\address{\it Department of Physics, School of Science, University of Tokyo, Tokyo~113-0033, Japan}
\date{June 15 1999}
\maketitle

\begin{abstract}
  We study the initial value problem of new inflation models in the
  scheme of supergravity. We find that an inflaton generally acquires
  a mass of the order of the Hubble parameter via gravitationally
  suppressed interactions with particles in the thermal bath of the
  universe. This additional mass term stabilizes dynamically the
  inflaton at the local maximum of the potential, which offers the
  initial condition of new inflation.
\end{abstract}


Inflation \cite{inf} is very attractive because it solves the horizon
and flatness problem of the standard big-bang cosmology and also gives
the primordial density fluctuations observed by the Cosmic Background
Explorer (COBE) satellite. In general inflation is caused by some
scalar field called the ``inflaton'' which has a very flat potential. 
Initially the inflaton is assumed to be displaced from the true
minimum of the potential and to give a nonzero vacuum energy. While
the inflaton rolls down toward the true minimum very slowly, this
vacuum energy dominates the energy of the universe and the
exponentially expanding era, i.e., inflation, is realized.

Here, we would like to consider inflation models in the
supersymmetric (SUSY) theories. The SUSY is the most attractive
extension beyond the standard model of particle physics, since it
stabilizes the electroweak scale against the radiative corrections and
realizes the unification of the standard gauge couplings.  Therefore,
it is important to investigate inflation in the scheme of SUSY and
also of its local version, i.e., supergravity.

So far, many SUSY inflation models have been proposed \cite{LR} and
they are roughly classified to three types: chaotic \cite{cha}, hybrid
\cite{hyb}, and new \cite{new} inflation. Within the framework of
supergravity new inflation is favored because it naturally gives
the solution to the gravitino problem. To avoid overproduction of
gravitinos, the sufficiently low reheating temperature $T_{R}
\lesssim 10^{8}$ GeV is required \cite{gra}. Such a low reheating
temperature is naturally offered in SUSY new inflation models
\cite{KMY,IY,DR}.

New inflation, however, suffers from a serious problem about the
initial value of the inflaton $\phi$ \cite{Linde}. Successful
inflation requires the initial value be very close to the local
maximum of the potential ($\phi \simeq 0$) not the true minimum ($\phi
= \langle \phi \rangle)$. But there is no reason to take such an
initial condition over the horizon scale, because the potential of the
inflaton is very flat to satisfy the slow roll conditions \cite{Linde}.
One might imagine that the inflaton is set to be $\phi = 0$ due to the
finite temperature effects via rapid interactions with particles in
the thermal bath. However, such interactions generally break the
flatness of the inflaton potential. Moreover, the inflaton usually
receives large fluctuations of thr order of the cosmic temperature so
that the required initial value of the inflaton is not realized over
the horizon scale.

Izawa, Kawasaki, and Yanagida \cite{IKY} found that another inflation
(called preinflation) which takes place before new inflation solves
the initial value problem.\footnote{Preinflations are also considered in
  a different context by Ref. \cite{Pre}.} During preinflation the
inflaton of new inflation acquires an additional mass term $\sim H_I^2
\phi^2$ ($H_I$ is the Hubble constant during preinflation) through the
supergravity effects. This effective mass is large enough for the
inflaton to oscillate around $\phi = 0$. Since the amplitude of the
oscillation decreases at the rate $R^{-3/2}$ ($R$ is the scale
factor), the inflaton rapidly settles at $\phi = 0$ if preinflation
lasts long enough. However, this preinflation is introduced besides
new inflation only to solve the initial value problem.

In this paper, we show that the inflaton, in general, gains the
effective mass of the order of the Hubble parameter in the
radiation-dominated universe through the gravitationally suppressed
interactions with particles in the thermal bath.  Therefore, the
inflaton rolls down toward $\phi =0$ dynamically, which explains the
initial condition of new inflation without introducing another
inflation.

Let us consider the K\"ahler potential for the inflaton in the
supergravity Lagrangian. In general, the K\"ahler potential for the
supermultiplets of the inflaton $\phi$ of new inflation and the
SUSY standard-model particles $\chi_i$ takes the form
\begin{eqnarray}
  K(\phi, \chi_{i}) &=&
  \left| \phi \right|^2 
  +
  \sum_i \left| \chi_i \right|^2 
  + \frac{k}{4} \left| \phi \right|^4
  + \sum_i \frac{k_i}{4} \left| \chi_i \right|^4
  \nonumber \\
  &&+
  \sum_{i} \lambda_{i}|\phi|^{2}|\chi_{i}|^{2} +
  \cdots \:,
  \label{KP}
\end{eqnarray}
where $k$, $k_i$, and $\lambda_i$ denote couplings of order unity and
the ellipsis denotes higher order terms. Here and hereafter, we set the
gravitational scale $M_{G} \simeq 2.4 \times 10^{18}$ GeV equal to
unity. The last term in Eq. (\ref{KP}) leads to the interaction
Lagrangian
\beq
  \CL_{\rm int} = \sum_{i} \lambda_{i} |\phi|^{2} 
           \del_{\mu}\chi_{i}^\ast \del^{\mu}\chi_{i} \:.
\label{eqn:int}
\eeq
Here we use the same letters for the corresponding scalar components of
the supermultiplets. Below the gravitational scale, these interactions
decouple and the inflaton cannot be thermalized. On the other hand,
the fields $\chi_i$ can be in thermal equilibrium, since they have
some charges of the standard-model gauge groups and also have some
Yukawa interactions.

We are now at the point to show how the interactions (\ref{eqn:int})
stabilize the inflaton at $\phi \simeq 0$, which is required for the
initial condition of new inflation. Just below the gravitational
scale, the inflaton is assumed to be located at the true minimum of
the potential ($\phi = \langle \phi \rangle$). For temperature $T
\simeq 10^{16}$--$10^{17}$ GeV, the interaction rates of $\chi_i$ via
gauge (or Yukawa) interactions become larger than the expansion of the
universe and they are in thermal equilibrium. After this epoch the
inflaton receives an additional mass through the interactions
(\ref{eqn:int}) as follows: The thermal average of the operators
$\del_{\mu}\chi_{i}^\ast \del^{\mu}\chi_{i}$ are estimated as
\begin{eqnarray}
    \langle \del_{\mu}\chi_{i}^\ast \del^{\mu}\chi_{i} \rangle &&
    = - \langle \chi_{i}^\ast \del_{\mu} \del^{\mu}\chi_{i} \rangle
    \nonumber \\
    &&
    = m_{\chi_i}^2(T) \langle \chi_{i}^\ast \chi_{i} \rangle
    = m_{\chi_i}^2(T) T^2 /12,
\end{eqnarray}
where $m_{\chi_i}^2(T)$ are thermal masses squared for $\chi_i$ and
given by $b_{\chi_i} T^2$ with $b_{\chi_i}$ being constants of order
unity depending on the interactions of $\chi_i$.  Therefore, the
inflaton obtains the effective mass of order of the Hubble parameter
as
\begin{eqnarray}
    m_{\rm eff}^2 = \sum_i \lambda_i b_{\chi_i} T^4 /12
    = c^2 H^2,
    \label{eqn:mass}
\end{eqnarray}
since $T^4 \sim H^2$ in the radiation dominated era. Here we assumed
$\lambda_i$ are positive so that the effective mass squared becomes
positive. Also, $c$ is a constant of order unity. This mass term
modifies the evolution of the inflaton, which is described by the
equation
\begin{eqnarray}
    \ddot{\phi} + 3 H \dot{\phi} + m_{\rm eff}^2 \phi = 0,
\end{eqnarray}
and we find $\phi \propto R^{ -1/2 + \sqrt{ 1/4 - c^2 } }$ using
$\dot{H} = -2H^{2}$ in the radiation-dominated universe. Then, for $c
> 1/2$, we obtain
\beq
  \phi = \phi_{\ast} \lmk \frac{R}{R_{\ast}} \rmk^{-\frac12} 
    \cos \lkk
           \sqrt{c^2 - \frac14} \ln \lmk \frac{R}{R_{\ast}} \rmk
         \rkk,
  \label{eqn:damped_osc}
\eeq
where $\phi_{\ast}$ and $R_{\ast}$ are the amplitude and scale factor
when the fields $\chi_{i}$ are in thermal equilibrium at temperature
$T_{\ast}$. Thus, for $T < T_{\ast}$, the inflaton oscillates around
$\phi = 0$ with amplitude decreasing as $R^{-1/2}$ and it lasts until
the vacuum energy $v^4$ of the inflaton becomes comparable to the
radiation energy. Thus, the initial value of the inflaton $\phi_{I}$
when new inflation starts becomes at most $(v/T_\ast)^{1/2} \langle
\phi \rangle$. If the factor $(v/T_\ast)^{1/2}$ is small enough, we
can avoid the initial value problem.\footnote{If we are allowed to
  tune the constant $c$, we can make the initial value $\phi_{I}$ much
  smaller than $\phi_{\ast}(v/T_{\ast})^{1/2}$. From Eq.
  (\ref{eqn:damped_osc}), $\phi_{I} = 0$ is achieved for $c = c_{0}$
  given by $c_{0}^{2}-1/4 = (\pi/2+m\pi)^{2}/\ln^{2}(T_{\ast}/v)$
  where $m$ is an integer.  Notice that due to the logarithmic
  dependence on $T_{\ast}/v$ of the period, $\phi_{I} \simeq 0$ is
  realized by mild tuning of $c(\simeq c_{0})$.} Note that the energy
density of the inflaton oscillation decreases as $R^{-5}$ so that the
energy is dominated by the radiation until the start of new inflation,
and also that the inflaton does not receive large fluctuations since
the interactions (\ref{eqn:int}) are suppressed by the gravitational
scale.

The present mechanism for the dynamical tuning of the initial value of
the inflaton applies for the generic new inflation models in
supergravity \cite{KMY,IY,DR}. As a concrete example, we consider a
model proposed by Izawa and Yanagida \cite{IY} with the superpotential
\begin{eqnarray}
    W = v^2 \Phi - \frac{g}{n+1} \Phi^{n+1},
\end{eqnarray}
where $v$ is the scale of the model, $g$ a coupling constant, and $n$
a positive integer with $n \geq 3$. Note that the coupling through the
superpotential between the inflaton and the standard SUSY particles
$\xi_{i}$ is prohibited by the $U(1)_{R}$ symmetry. From the K\"ahler
potential (\ref{KP}) the inflaton $\phi = \sqrt{2}({\rm Re} \Phi)$
obtains the potential as
\begin{eqnarray}
    V(\phi) = v^4 - \frac12 k v^4 \phi^2
    - \frac{ g }{ 2^{\frac{n}{2}-1} } v^2 \phi^n
    + \frac{ g^2 }{ 2^n } \phi^{2n},
\end{eqnarray}
for $\phi < \langle \phi \rangle = \sqrt{2} \langle \Phi
\rangle$. The true vacuum of the inflaton is estimated as
\begin{eqnarray}
    \langle \Phi \rangle 
    = \langle \phi \rangle / \sqrt{2} 
    = \left( v^2/g \right)^{1/n}.
\end{eqnarray}
It has been shown in Ref. \cite{IY} that the slow-roll conditions for
the inflaton are satisfied for $k < 1$ and $\phi \lesssim
\phi_f$ where
\begin{eqnarray}
    \phi_f \simeq \sqrt{2} 
    \left( \frac{ (1-k) v^2 }{ gn (n-1) } \right)^{\frac{1}{n-2}},
\end{eqnarray}
which provides the value of $\phi$ when inflation ends. The field
value $\phi_N$, when the observable universe crossed the horizon
during inflation, is given by
\begin{eqnarray}
    \phi_N \simeq \sqrt{2} 
    \left( \frac{ v^2 k }{ n g } \right)^{\frac{1}{n-2}}
    \exp \lkk - k \lmk N + 
        \displaystyle{
        \frac{ k n - 1 }{ k(1-k)(n-2) }} \rmk \rkk,
\end{eqnarray}
with $N \sim 50$. Also, the amplitude of the density fluctuations
$\delta \rho /\rho$ is given by
\begin{eqnarray}
    \frac{ \delta \rho }{\rho} 
    \simeq \frac{ 1 }{ 5 \sqrt{3} \pi }
    \frac{ V^{3/2}(\phi_N) }{ \left| V'(\phi_N) \right| }
    \simeq\frac{ 1 }{ 5 \sqrt{3} \pi }
    \frac{ v^2 }{ k \phi_N },
\end{eqnarray}
and the spectral index $n_s$ of the density fluctuations is given by
$n_s \simeq 1 - 2 k$. Since the COBE data \cite{COBE} implies $n_s =
1.0 \pm 0.2$, we should take $k \lesssim 0.1$, and the COBE
normalization gives $v^2/ ( k \phi_N ) \simeq 5.3 \times 10^{-4}$. For
example, we consider the case of $n = 6$ with $g=0.1$ and $k=0.02$.
Then, we obtain $v \simeq 4.9 \times 10^{14}$ GeV, $\la \phi \ra
\simeq 2.9 \times 10^{17}$ GeV, and $\phi_N \simeq 9.6 \times 10^{15}$
GeV. In the present scenario, the initial value of the inflaton
$\phi_I$ is given by $\phi_I \lesssim ( v/T_\ast )^{1/2} \langle \phi
\rangle$ due to the additional mass (\ref{eqn:mass}) so that it becomes
smaller than $\phi_N$ if $T_\ast \gtilde 1.6 \times
10^{17}$\,GeV.~\footnote{ If one takes smaller $g$ or larger $n$, the
  required $T_\ast$ becomes smaller. If one takes larger $k$, the
  required $T_\ast$ becomes larger, but in this case the spectrum of
  the density fluctuations deviates from the scale invariant form too
  much. }  Therefore, the existence of the nonrenormalizable
interactions (\ref{eqn:int}),\footnote{Through these interactions, the
  inflaton can decay into the SUSY standard-model particles and the
  universe reheats. In the present model, the reheating temperature
  $T_{\rm RH}$ becomes $T_{\rm RH} \sim 10^{6-8}$ GeV, which is low
  enough to avoid the overproduction of gravitinos \cite{gra}.} which
is inevitable in the framework of the supergravity, can naturally
solve the initial value problem of new inflation.

Finally, we would like to remark on the flatness problem in the
present model. Since the energy scale at the beginning of new
inflation is much lower than the gravitational scale, the question may
be raised why the universe lives so long (flatness problem). The
preinflation, if it occurs near the gravitational scale, may avoid it
\cite{IKY}. However, this difficulty does not occur if the universe
was created quantum mechanically as flat, or open.

This work was partially supported by the Japanese Society for the
Promotion of Science (T.A.,M.Y.) and ``Priority Area: Supersymmetry and
Unified Theory of Elementary Particles(\#707)''(M.K.).


\begin{references}

\bib{inf}A. H. Guth, \PRD{23}{347}{81},
K. Sato, \MNRAS{195}{467}{81}.

\bib{LR}See, for a review, D. H. Lyth and A. Riotto,
\PRT{314}{1}{99}.

\bib{cha}A. D. Linde, \PLBold{129}{177}{83}.

\bib{hyb}A. D. Linde, \PLB{259}{38}{91}.

\bib{new}A. D. Linde, \PLBold{108}{389}{82},
A. Albrecht and P. J. Steinhardt, \PRL{48}{1220}{82}.

\bib{gra} M. Yu. Khlopov and A.D. Linde, \PLBold{138}{265}{84}; J.
Ellis, E. Kim and D.V. Nanopoulos, {\it ibid}, {\bf 145B}, 181(1984);
M. Kawasaki and T. Moroi, \PTP{93}{879}{95}.

\bib{KMY}K. Kumekawa, T. Moroi, and T. Yanagida,
\PTP{92}{437}{94}.

\bib{IY}K. I. Izawa and T. Yanagida,
\PLB{393}{331}{97}.

\bib{DR}M. Dine and A. Riotto,
\PRL{79}{2632}{97}.

\bib{Linde}For example, A.D. Linde, Particle Physics and Inflationary
Cosmology, (Harwood, Chur, Switzerland, 1990).
 
\bib{IKY}K. I. Izawa, M. Kawasaki, and T. Yanagida,
\PLB{411}{249}{97}.

\bib{Pre}J. A. Adams, G. G. Ross, and S. Sarkar,
\NPB{503}{405}{97};
G. Lazarides and N. Tetradis,
\PRD{58}{123502}{98};
C. Panagiotakopoulos and N. Tetradis,
\IBD{59}{083502}{99}.
 
\bibitem{COBE}
  C.L. Bennett et al.,
  Astrophys. J. Lett. {\bf 464},L1 (1996).
\end{references}
\end{document}